\documentclass[prd,showpacs,nobibnotes,twocolumn,nobalancelastpage]{revtex4}

\usepackage{bm}
\usepackage{graphicx}
\usepackage{epsf}

\def\mpc{\,h^{-1}{\rm Mpc}}
\def\kpc{\,h^{-1}{\rm kpc}}

\def\msun{\,h^{-1}{\rm M}_\odot}

\makeatletter

\newcommand{\Rmnum}[1]{\expandafter\@slowromancap\romannumeral #1@}
\makeatother

\newcommand{\mnras}{MNRAS}
\newcommand{\apjl}{ApJ}

\newcommand{\jcap}{JCAP}
\newcommand{\araa}{ARA\&A}

\begin{document}


\title{Nonlinearities in modified gravity cosmology --- \Rmnum{2}. Impacts 
of modified gravity on the halo properties}

\author{Youcai Zhang$^1$}
\email{yczhang@shao.ac.cn}
\author{Pengjie Zhang$^1$}
\author{Xiaohu Yang$^1$}
\author{Weiguang Cui$^2$}

\affiliation{$^1$Key Laboratory for Research in Galaxies and Cosmology,
  Shanghai Astronomical Observatory; Nandan Road 80,
  Shanghai 200030, China; E-mail: yczhang@shao.ac.cn}

\affiliation{$^2$Astronomy Unit, Department of Physics, University of
Trieste, via Tiepolo 11, I-34131 Trieste, Italy}

\begin{abstract}
The statistics of dark matter halos is an essential component of understanding
the nonlinear evolution in  modified  gravity  cosmology.  Based  on  a series  of
modified gravity  N-body simulations, we  investigate the halo  mass function,
concentration  and  bias.  We model the impact of modified gravity by a single
parameter $\zeta$, which  determines the enhancement 
of  particle  acceleration  with  respect  to GR,  given  the  identical  mass
distribution ($\zeta=1$    in    GR).   We select snapshot redshifts such that
the  linear matter power spectra of different gravity models are identical, in order to isolate
the impact of gravity beyond modifying the linear growth rate. At the baseline
redshift corresponding to $z_S=1.2$ in the standard $\Lambda$CDM,  for    a    $10\%$   deviation    from
GR($|\zeta-1|=0.1$),  the measured  halo  mass function  can  differ by  about
$5-10\%$, the  halo concentration by  about $10-20\%$, while the halo  bias
differs significantly less.  These results demonstrate that  the halo
mass  function and/or  the halo  concentration are  sensitive to  the nature
of gravity and may be used to make interesting constraints along this line.
\end{abstract}


\maketitle
\section{INTRODUCTION}\label{sec_intro}

Dark  matter halos  are  prominent  structures in  the  dark universe.   Their
abundance,  density  profile  and  clustering  (halo  bias)  contain  valuable
cosmological information. In particular, dark matter halos form and grow under
gravitational instability.  Hence halo  properties contain rich information on
the nature of  gravity at $\sim$ Mpc scales and above,  and can provide strong
gravity  constraints \citep{Narikawa2012,Thomas2011a,Thomas2011b,Sartoris2011,
  LiB2011, Lombriser2011, Mak2011, LiYin2011, Jain2011, LiMiao2011, Allen2011,
  Clifton2011,  Ferraro2011, Schmidt2009a,  Schmidt2009c,  ZhaoG2011, Jain10}.
These  properties  also provide  a  powerful  tool  to understand  the  matter
clustering  through the  halo model  \cite{Jing98,Cooray02}, which  is  also a
sensitive measure of gravity.

In  \cite{Cui10}  (hereafter  paper I)  we  run  a  controlled set  of  N-body
simulations with identical  initial condition to study the  impact of modified
gravity  (MG) on  the matter  power spectrum.   These simulations  adopt  a MG
parameterization  with a single  parameter $\zeta$.   $\zeta$ is  the relative
enhancement of  nonrelativistic particle acceleration with  respect to general
relativity  (GR),  given  the  identical  mass  distribution.  More
specifically, it is
the quantity that enters into the $\psi$-$\rho$ relation in Fourier space, 
\begin{equation}
\label{eqn:psi}
k^2\psi=-\zeta4\pi G a^2\delta\rho\ .  
\end{equation}
Here,  $\delta \rho\equiv \rho-\bar{\rho}$ and $\rho$ is
the matter density. $\psi$ is defined through $ds^2=-(1+2\psi)dt^2+a^2(1+2\phi)d{\bf
  x}^2$. $\psi$ is the only gravitational potential that nonrelativistic
particles can sense.  So the value of $\zeta(k,z)$ as a function of
scale and redshift, along with the initial condition and the expansion
rate of the universe, completely fixes the evolution of the matter
density and velocity fields. This is the major reason that we adopt this single
parameter parametrization on modified gravity N-body simulations. For more detailed
discussion on this parametrization, please refer to paper I. 

Neverthless, such parameterization is highly simplified in the sense that
it  lacks any  screening mechanism  \cite{Jain10} required  to pass  the solar
system  tests. For  this  reason,  these simulations  lose  the capability  to
explore  the rich  consequences induced  by the  MG  environmental dependence.
These features  have been explored in  advanced simulations on  $f(R)$ and DGP
\cite{Schmidt2009a,  Oyaizu08a, Oyaizu08b,  ZhaoGB2011,  Schmidt2009b, Chan09,
  Li12}.  Nevertheless,  our simulations benefit from costing  no extra time
than the ordinary CDM simulations.  So in principle one can run a large number
of such simulations within reasonable amount of time to fairly sample relevant
parameter space.  One then interpolates/extrapolates the simulation results to
explore  the  whole relevant  parameter  space  and  understand the  nonlinear
evolution for  $\zeta$ of  general spatial and  time dependence.  The  hope is
that, by  choose appropriate spatial and  time dependence in  $\zeta$, one can
effectively take the  environmental dependence into account \footnote{Strictly
  speaking,  MG described by  a spatial  and time  dependent $\zeta$  is still
  environmental  independent.  Nevertheless,  $\zeta$  with spatial  and  time
  dependence is promising to mimic  the statistically averaged properties of a
  environmental  dependent  MG model.  In  such  case,  the interpretation  of
  $\zeta$ may not be straightforward.}.

In the current paper, we will  examine the halo properties in these simplified
MG  simulations, including  the halo  abundance, concentration  and bias  as a
function  of  mass and  redshift.   In  the  standard framework  of  structure
formation in GR, these properties,  in particular the halo abundance and bias,
are largely  fixed by  the linear matter  power spectrum at  the corresponding
epoch. We utilize  this property to better isolate the impact  of MG.  We will
compare the  above properties not  at the same  redshift, but at  redshifts of
identical linear  matter power spectrum.  The  same method is  also adopted in
paper  I. It  has  extra benefit  of  reducing cosmic  variance and  numerical
artifact, since we  focus on the differences between  different MG simulations
with identical initial condition.

This paper is organized as follows.  We briefly describe our simulations in \S
\ref{sec_data}  and present results  in \S  \ref{sec_imethod}. We  discuss and
summarize in \S \ref{sec:conclusion}.

\section{N-body Simulation}\label{sec_data}

We  run  a  set  of   simulations  with  the  TreePM  parallel  code  GADGET-2
\citep{Springel2005} at Shanghai Astronomical Observatory. All the simulations
evolved $512^3$  dark matter  particles in  a periodic box  of $300\mpc$  on a
side. The cosmological parameters used in the simulations are $\Omega_{\rm m}=
\Omega_{\rm  dm}+\Omega_{\rm b}  =0.276$,  $\Omega_{\rm b}=0.046$,  $h=0.703$,
$\Omega_\Lambda=0.724$,  $n=1$, and  $\sigma_8=0.811$. The  particle  mass and
softening   length   are   $1.541\times10^{10}   \msun$  and   $12.89   \kpc$,
respectively.   Glass-like cosmological initial  conditions were  generated at
redshift $z=100$ using the Zel'dovich approximation.

In  this  paper  as  in   Paper  \Rmnum{1},  the  modified  gravity  model  is
characterized by a single  parameter $\zeta$, which determines the enhancement
of  particle  acceleration  with  respect  to GR,  given  the  identical  mass
distribution  ($\zeta=1$ in  GR). All  the simulations  for  different $\zeta$
values have the same linear power spectrum, such that
\begin{equation}
P_L(k;z_S, \zeta=1) = P_L(k;z_\zeta, \zeta),
\end{equation}
where $P_L$  is linear matter power  spectrum, $z_S$ and  $z_\zeta$ denote the
redshifts   in    the   standard   $\Lambda$CDM   and    the   MG   cosmology,
respectively.  Since all  the  simulations begin  with  the identical  initial
condition, the above relation means that
\begin{equation}
\label{eqn:D_z}
D(z_S, \zeta=1) = D(z_\zeta, \zeta),
\end{equation}
where $D(z,\zeta)$ is the linear density growth factor.  Given redshifts $z_S$
in  the  $\Lambda$CDM,  through   Eq.~(\ref{eqn:D_z})  we  can  calculate  the
corresponding redshifts $z_\zeta$, which are shown in Table \Rmnum{1} of Paper
\Rmnum{1}. In this paper,  we mainly focus on the redshift $z_S  = 1.2$ in the
$\Lambda$CDM and the  corresponding redshifts in the MG  universe (10th row of
Table \Rmnum{1} in Paper \Rmnum{1}).

\begin{figure}
\includegraphics[width=0.45\textwidth]{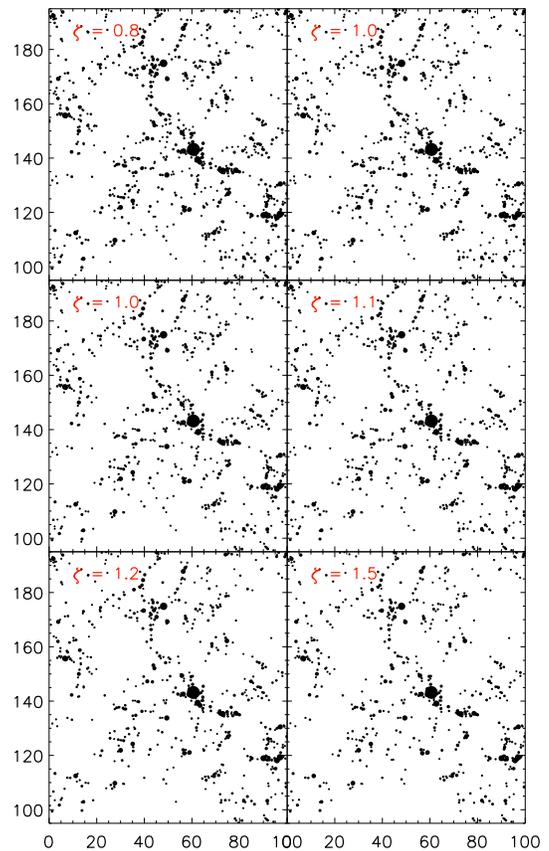}
\caption{Comparison  of   halo  distributions  from   N-body  simulations  for
  different $\zeta$ values  at the baseline redshift $z_S=1.2$,  which is that
  of $\zeta=1$ ($\Lambda$CDM). The panels  show the dark matter halos with the
  mass  $M_{200}\ge10^{11.5}\msun$  in  slices  $100\times100\times8h^{-1}{\rm
    Mpc}$, and the  sizes of the dots are proportional  to the radii $R_{200}$
  of the halos.}
\label{fig:slice}
\end{figure}

Dark matter halos  were identified from the simulation  at each snapshot using
the  standard  friends-of-friends   (FOF)  algorithm\citep{Davis1985}  with  a
linking  length  of $b=0.2$  times  the  mean  interparticle separation.   For
currently favored cosmologies, $15$-$20\%$ of $b=0.2$ FOF halos have irregular
substructure   or   have   two   or   more  major   halo   components   linked
together\citep{Lukic2009}. Only  isolated, relaxed  halos are well-fit  by the
Navarro-Frenk-White  (NFW) profile\citep{Navarro1997},  therefore, we  use the
virial  halo mass  to fit  NFW  profile in  the simulations.  The mass
$M_{200}$ is defined as the interior  mass within a sphere of radius $R_{200}$
centered on the most bound particle of the halo, where $R_{200}$ is the radius
within  which the over-density  is $\Delta_{200}=200$  times the  mean density
$\rho_{\rm mean}$  of the universe.  Thus, the mass  and radius of a  halo are
related by
\begin{equation}
M_{200} = \frac{4}{3}\pi \Delta_{200} \rho_{\rm mean} R^3_{200},
\end{equation}
where  $\rho_{\rm mean}$  is  the mean  density  of the  universe at  redshift
$z$.   Note  that some authors use  a value $\Delta_{\rm vir}$  motivated by the
spherical collapse  model, where  the corresponding
radius is specified as $R_{\rm vir}$. This value is both redshift and gravity
dependent. It gives  $\Delta_{\rm vir}=339.5$ at  $z=0$ and $178$ at $z\gg 1$ for
our  adopted $\Lambda$CDM  parameters \citep{Eke1996}. It also shows significant
deviation in MG models \cite{Schmidt2009a}. Since  we  will compare  the  halo
properties of different MG models  at 
different redshifts  (e.g.  $z_s$ vs. $z_\zeta$), this latter  choice could mix
the  impact of  background  cosmology with  the  impact of  gravity. For  this
reason, we will not adopt this latter definition.  Unless otherwise specified,
throughout the  paper we  use a  fixed value $\Delta=200$  to define  the halo
mass. 

Fig.~\ref{fig:slice}  shows the distributions  of dark  matter halos  with the
mass $M_{200}\ge10^{11.5}\msun$ in a slice of thickness $8h^{-1}{\rm Mpc}$ for
different $\zeta$ values at the  baseline redshift $z_S=1.2$, which is that of
$\zeta=1$  ($\Lambda$CDM).   According  to  visual   inspection,  the  general
appearances of the large-scale structures  are remarkably similar, as a result
of the same linear power  spectrum for different $\zeta$ values.  Although the
shape of Cosmic Web (the pattern on clusters, filaments, sheets, and voids) is
more  or  less  preserved  in  the  $\zeta\ne1$  runs  when  compared  to  the
$\Lambda$CDM case,  we can  depict arising differences  when looking  into the
internal properties of dark matter halos.

\section{Numerical Results}
\label{sec_imethod}

In this  section we show the halo  mass function, concentration and  bias as a
function of $M_{200}$.  Structure grows more slowly in $\zeta<1$ models than
that in GR. So for too low $z_S$ in $\Lambda$CDM, structure growth in a
$\zeta<1$ universe may not be
able to catch up with that of $\Lambda$CDM even at $z_\zeta=0$ (table I, paper
I).  For this
reason, we only compare the results corresponding to $z_S=1.2$
in $\Lambda$CDM, for which we can compare the halo statistics for all
$\zeta$.

\subsection{The halo mass function}

\begin{figure}
\includegraphics[width=0.4\textwidth]{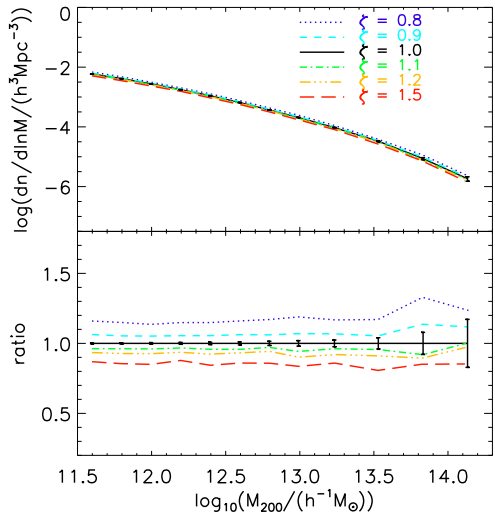}
\caption{Measured halo  mass functions  from N-body simulations  for different
  $\zeta$ values at redshift $z_\zeta$  where the linear matter power spectrum
  is identical  to that in  $\Lambda$CDM ($\zeta=1$) at the  baseline redshift
  $z_S=1.2$.  Since  all simulations begin  with the identical  condition, the
  measured halo  mass function has similar  Poisson error. Hence  we only show
  the  Poisson  error bars  of  $\Lambda$CDM  to  demonstrate the  measurement
  uncertainty.  The lower panel shows the ratio of the measured mass functions
  of MG models to the one in $\Lambda$CDM when they have the same linear power
  spectrum.  The plotted errors  are the  same Poisson  errors showing  in the
  upper panel. Since  the Poisson errors of different  simulations have strong
  positive correlations,  they largely  cancel when taking  the ratio.  So the
  plotted  errors  represent a  conservative  upper  limit  of the  halo  mass
  function ratio. The lower panel then shows that the observed deviations from
  a universal mass function are significant.  }
\label{fig:massfunction}
\end{figure}

One of  the long-standing efforts in  precision cosmology is  to determine the
mass function of dark matter halos ${\rm d}n/{\rm d}M$, which is the number of
halos per unit volume per unit mass. It is a sensitive probe of gravity and
has been used to put strong constraints on modified gravity models
(e.g. \cite{Schmidt2009b}  on $f(R)$ gravity).

Although the number density of halos of a
given  mass  depends  on  the  shape  and amplitude  of  the  power  spectrum,
analytical work  has suggested that the  halo abundance can be  expressed by a
universal   functional   form   when    expressed   in   terms   of   suitable
variables \citep{Press1974, Bond1991, Sheth1999}. A convenient form to describe
halo abundance can be expressed as \citep{Jenkins2001}
\begin{equation}\label{eqn:ms}
\frac {{\rm d}n} {{\rm d}M}(M,z) = f(\sigma) \frac {\rho_0}{M}
\frac {{\rm d~ln}[\sigma^{-1}(M,z)]}{{\rm d}M},
\end{equation}
where  $\rho_0$  is the  background  matter  density  at redshift  $z=0$,  and
$\sigma^2(M,z)$ is  the variance  of the linear  matter power spectrum  over a
length $R$,
\begin{equation}\label{eqn:gf}
\sigma^2(M,z) = \frac{D^2(z)}{2\pi^2} \int_0^\infty k^2P(k)W^2(kR(M))dk,
\end{equation}
where  $W(x)=3[\sin(x)-x\cos(x)]/x^3$  is the  Fourier  transformation of  the
top-hat filter, $R(M)=(3M/4\pi \rho_0)^{1/3}$ is the smoothed radius with halo
mass $M$,  and $D(z)$ is the  growth factor. 

This definition  of mass function has  been  widely  examined  against  N-body
simulations  and  useful  fitting 
functions $f(\sigma)$ have been provided by several authors \citep{Jenkins2001,
  Warren2006,   Tinker2008,  Bhatt2010}.   Nevertheless, deviations from the
universality have been reported. For example,  \citet{Bhatt2010}  found
through their N-body simulations that,  for their $\omega$CDM
cosmological  models  (where  the  dark  energy equation  of  state  parameter
$\omega$ is constant in time, but $\omega\ne-1$), the universality of the mass
function is  systematically broken at a level of  $10\%$.  Deviations from the
universality are also detected in interacting dark energy
models \citet{Cui2012}. \citet{Cui2012} reported that deviations can exceed 
$\sim10\%$ for most  of the models in the  high mass end.

Fig.~\ref{fig:massfunction} shows  the measured  halo mass functions  from our
simulations for  different $\zeta$ values at the  baseline redshift $z_S=1.2$,
which is that of $\zeta =1$ ($\Lambda$CDM). The $M_{200}$ halo mass is used in
the plot in  order to get consistent comparisons  with concentration and bias,
where the same mass definition are adopted.  We have checked our measured halo
mass functions for $\Lambda$CDM simulation,  which agree well with the fitting
function proposed by \cite{Sheth2001}.

In order to  assess the difference, the Poisson error bars  are only added for
$\zeta=1$. Because of the same initial condition been used, the error bars for
$\zeta\ne1$ have the similar  forms. In Fig.~\ref{fig:massfunction}, the lower
panel shows the ratio  of the measured mass functions of MG  models to the one
in $\Lambda$CDM. The Poisson error bars  for $\zeta=1$ are added to assess the
maximum  limit.   Since  $D(z_\zeta,\zeta)=D(z_S,\zeta=1)$, the  linear  power
spectrum and hence  $\sigma(M,z)$ are identical. If the  halo mass function is
indeed universal  and described by  Eq.~(\ref{eqn:ms}) and Eq.~(\ref{eqn:gf}),
the mass function in the MG model  should be the same as that in $\Lambda$CDM.
However,  the result  in Fig.~\ref{fig:massfunction}  shows the  difference of
mass  functions  between $\zeta=0.8$  and  $\zeta=1.0$($\Lambda$CDM) is  about
$20\%$  with the halo  mass range  from $10^{11.5}\msun$  to $10^{14.0}\msun$.
The difference between  $\zeta=1.2$ and $\zeta=1.0$ is about  $10\%$.  We find
that  the  difference of  $\zeta=1.2$  is  about  twice that  of  $\zeta=0.8$,
although they have the same $20\%$  deviation from GR.  For a $10\%$ deviation
from GR($\zeta=0.9$,  or $\zeta=1.1$), the measured mass  functions can differ
by about $5-10\%$.  Interestingly, the difference of $\zeta=0.9$ is also about
twice that of $\zeta=1.1$.  In general, for the identical deviation of $\zeta$
from unity,  the deviation of halo  mass function of $\zeta<1$  is about twice
that of $\zeta>1$, which implies that the decrease of gravity can lead to more
difference of  halo mass  function.  This  behavior can also  be found  in the
bottom-right panel  of Fig.~$4$ in Paper  \Rmnum{1} for the  comparison of the
nonlinear power  spectra. For  the same deviation  of $\zeta$ from  unity, the
deviation  of nonlinear power  spectra of  $\zeta<1$ are  about twice  that of
$\zeta>1$ at $\Delta^2\sim10$.

If the halo  abundance is completely determined by the  shape and amplitude of
the  linear power  spectrum, the  halo  mass functions  for different  $\zeta$
simulation should  have the  identical form.  However,  based on  the measured
halo mass  functions from N-body  simulations for different $\zeta$  values at
the same linear power spectrum, we find that this is not the case, which means
that the halo  abundance also depends on the structure  growth history and the
underlying  gravity. The  deviation of  halo abundance  becomes larger  if the
deviation  of  $\zeta$  from unity  is  larger.   The  halo mass  function  is
therefore a sensitive probe for our modified gravity model.

\subsection{Halo concentration}

\begin{figure}
\includegraphics[width=0.4\textwidth]{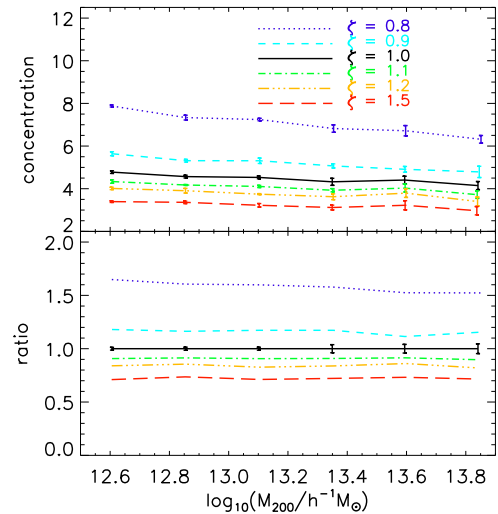}
\caption{Measured  halo concentrations from  N-body simulations  for different
  $\zeta$ values  at the baseline  redshift $z_S=1.2$.  The lower  panel shows
  the ratio  of the measured  halo concentrations of  MG models to the  one in
  $\Lambda$CDM.  For  clarity,  we  plot  the  error  bars  of  the  $\zeta=1$
  result.  As  explained  in  Fig. \ref{fig:massfunction},  these  errors  are
  conservative  upper  limit.  Hence  the  observed  differences  in  $c$  are
  statistically significant.  Difference in  the nature of gravity contributes
  to these  differences. However, since  $z_S\neq z_\zeta$, difference  in $c$
  also  arises  from  difference  in  the  background  matter  density.   More
  systematical studies  are required to disentangle  the two and  to infer the
  nature of gravity from the halo concentration. }
\label{fig:concentration}
\end{figure}

The  concentration parameter  has been  widely used  to describe  the internal
structure of halos\citep{Jing2000, Bullock2001, Zhao2003, Neto2007, Zhao2009}.
Recently,  \citet{Jeeson2011}   investigated  the  correlation   between  nine
different dark matter  halo properties, and they claimed  that while the scale
of  a halo  is set  by its  mass, the  concentration is  the  most fundamental
property.   In this  paper,  the NFW  form is  used  to fit  the halo  density
profiles,  which  can  be approximated  by  a  simple  formula with  two  free
parameters,
\begin{equation}
\label{eqn:nfw}
\frac {\rho(r)}{\rho_{\rm mean}} = \frac {\delta_{\rm c}}
{(r/r_s)(1+r/r_s)^2},
\end{equation}
where $r_s$ is a scale radius and $\delta_{\rm c}$ is a characteristic density
contrast.  The concentration  of a  halo is  defined as  $c=R_{200}/r_s$, thus
$\delta_{\rm c}$ and concentration are linked by the relation
\begin{equation}
\label{deltac}
\delta_{\rm c}=\frac {200}{3} \frac{c^3}{[{\rm ln}(1+c)
-c(1+c)]}.
\end{equation}
In our calculation of concentration from  the simulation, the radius of a halo
is divided in uniform logarithmic bins ($\Delta \log_{10}=0.1$), starting from
the  radius  where  the  bin  contains  at least  $20$  particles  to  $R_{\rm
  max}=R_{200}$ (or to $R_{\rm vir}$ in case halo masses are defined using the
over-density    $\Delta_{\rm    vir}$).     Due    to    the    relation    of
equation~\ref{deltac}, at  given halo mass  $M_{200}$, there is a  single free
parameter   in  equation~\ref{eqn:nfw},   which  can   be  expressed   as  the
concentration  parameter $c$.   The  best-fit concentration  parameter can  be
computed from simulation by minimizing  the rms deviation $\sigma$ between the
binned $\rho(r)$ and the NFW profile,
\begin{equation}
\sigma^2 = \frac{1}{N_{\rm bins}} \sum_{i=1}^{N_{\rm bins}}
[\log_{10} \rho_i - \log_{10} \rho_{\rm NFW}(c)]^2\,.
\end{equation}

Fig.~\ref{fig:concentration}  shows the  measured concentration  of  the halos
contained at least $200$ particles for different $\zeta$ values.  In each mass
bin,  the  halos are  randomly  divided into  five  parts.   Then the  average
concentration  of  each  part  is  calculated.   The  error  bars  plotted  in
Fig.~\ref{fig:concentration}  is  the   standard  deviation  of  five  average
concentrations.  In  general the errors  of the average concentrations  of the
total halos can  be reduced by a factor of $\sqrt{5}$.  For $\zeta=1$, we find
that    the    concentrations     of    halos    with    $N>1000$    particles
($M_{200}>13.18\msun$) can  be approximately fitted  by a power  law $c\propto
M_{200}^{-0.1}$,   which  is   in   good  agreement   with   the  results   of
\citet{Neto2007}.  For  the halos with  $N<1000$ particles, we find  the power
law index  is less than  $-0.1$, which is  in good agreement with  the results
calculated  by  the model  of  \citet{Zhao2009}  \footnote{A calculator  which
  allows one to  interactively generate data for any  given cosmological model
  is provided at http://www.shao.ac.cn/dhzhao/mandc.html.}.

Because  the  main  focus in  this  paper  is  to  compare the  difference  of
concentration   for  different  $\zeta$   values,  in   the  lower   panel  of
Fig.~\ref{fig:concentration}, the ratio of the measured halo concentrations is
plotted.  For a  $10\%$ deviation from GR ($|\zeta-1|<0.1$),  the deviation of
halo  concentration can  differ by  $10-20\%$, which  means that  the modified
gravity can  significantly affect the  structure formation on small  scale and
the  internal properties  of  the  dark matter  halos.   For $\zeta=0.8$,  the
deviation of halo concentration even  differ by larger than $50\%$ compared to
that  of $\zeta=1.0$.   Here again,  with similar  difference in  $\zeta$, the
concentration difference of $\zeta< 1$ is much larger than that of $\zeta>1$.

The  discrepancy between the  different $\zeta$  values is  mainly due  to the
different  background  density  although  they  have the  same  linear  matter
perturbation. Since structure grows  faster in $\zeta>1$ cosmology ($z_\zeta >
z_S$), halos  in this universe form  in a background with  higher mean density
($\propto(1+z)^3$).    We   then    expect    them   to    have   a    smaller
concentration\citep{Zhao2009}.  For  the same  reason,  we  expect halos  with
$\zeta<1$ have a larger concentration.  The behavior of the concentrations for
different $\zeta$ values is roughly similar to the trend of the concentrations
corresponding to  the redshift $z_\zeta$ in $\Lambda$CDM  models. Large values
of the ratio of the concentration  parameters from unity implies that the halo
concentration  is a  valuable property  to detect  different  modified gravity
models.

\subsection{Halo bias}

\begin{figure}
\includegraphics[width=0.4\textwidth]{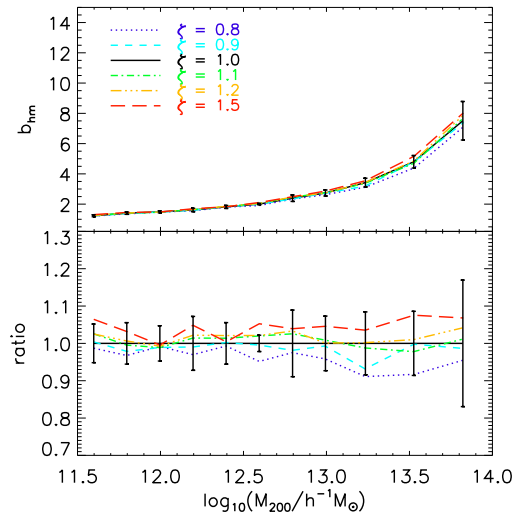}
\caption{Measured  halo bias  from  N-body simulations  for different  $\zeta$
  values at the baseline redshift  $z_S=1.2$.  The lower panel shows the ratio
  of the  measured halo bias of  MG models to  the one in $\Lambda$CDM.  As in
  previous  plots,  the errors  are  that  of  $\Lambda$CDM. As  explained  in
  previous plots, these errors are  conservative upper limit. For this reason,
  the  observed difference  in  the halo  bias  is statistically  significant,
  despite the apparently large error bars. }
\label{fig:bias}
\end{figure}

We  define the  bias  of dark  matter  halos as  the ratio  of  the halo  mass
cross-power spectrum to the dark matter power spectrum
\begin{equation}
b(k,M_{200}) = \frac {P_{hm}(k,M_{200})}{P_{mm}(k)},
\end{equation}
where  $P_{hm}(k,M)$  denotes the  cross-power  spectrum  with  halos of  mass
$M_{200}$, and  $P_{mm}(k)$ is  the dark matter  power spectrum.  This measure
dose not require a shot-noise correction, and it yields better statistics when
the halos  become sparse.  The power spectrum  is calculated  using Daubechies
wavelet mass assignment, which avoids  the sampling effect. Although halo bias
is scale dependent in the quasi-linear  and nonlinear regime, here we focus on
the large-scale bias, where $b$ is independent of $k$. We calculate $b$ as the
average over the $5$ largest  wavelength modes ($k\lesssim0.1 h/{\rm Mpc}$) in
the simulation. We also check these results against bias as defined by $b_{hh}
= \sqrt {P_{hh}/P_{mm}}$,  and we find that there is  a good agreement between
$b_{hm}$ and $b_{hh}$.

In the $\Lambda$CDM cosmological model,  the linear halo bias can be expressed
as a function of $\nu$, \citep{Sheth2001,Mo1996,Seljak2004, Tinker2010}, where
$\nu=\delta_c/\sigma$ is  the ratio of the critical  over-density required for
collapse to the rms density fluctuation.

Fig.~\ref{fig:bias}  shows measured  halo biases  from N-body  simulations for
different $\zeta$ values  at the baseline redshift $z_S=1.2$.   The error bars
is the standard deviation of the  $5$ largest wavelength modes. The error bars
for different $\zeta$ values have  very similar amplitude and only $\zeta=1.0$
has  been plotted  in  the figure.  The lower  panel  shows the  ratio of  the
measured halo bias of  MG models to the one in $\Lambda$CDM.  We find that the
difference of  halo bias  for different $\zeta$  values becomes larger  as the
increase of  halo mass.  However, the  deviations of halo bias  differ by less
than $10\%$  from GR ($\zeta=1$)  for all the  $\zeta\ne1$ values used  in our
simulation.  This behavior  is expected due to the halo bias  is a function of
$\nu$ in $\Lambda$CDM model. The simulations for different $\zeta$ values have
the  same linear  power spectrum  and growth  factor, therefore,  according to
Eq.~(\ref{eqn:gf}) they  have the  same density fluctuation  $\sigma$. Besides
$\delta_c$ is  weakly dependent  on the redshift,  thus the difference  of the
halo bias for different $\zeta$ values is very small. We can conclude that the
halo bias is a weak statistics to detect the modified gravity model.

\begin{figure}
\includegraphics[width=0.45\textwidth]{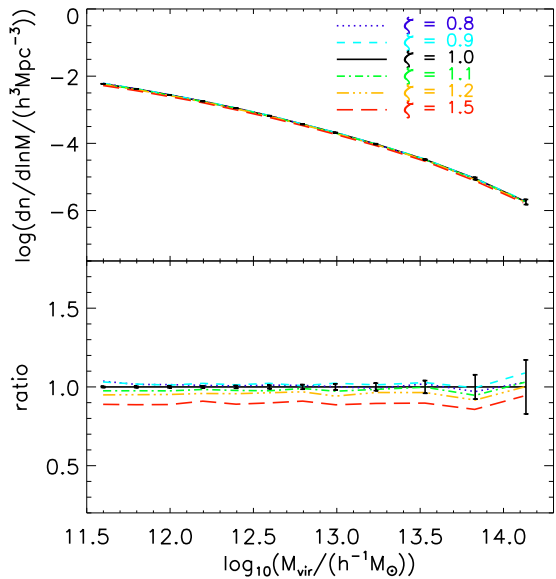}
\caption{Same  as Fig.~\ref{fig:massfunction}  but  for the  virial halo  mass
  $M_{\rm  vir}$ defined  by a  value  $\Delta_{\rm vir}$  from the  spherical
  collapse model.  }
\label{fig:ms_vir}
\end{figure}

\section{Discussion and Conclusion}
\label{sec:conclusion}

\begin{figure}
\includegraphics[width=0.45\textwidth]{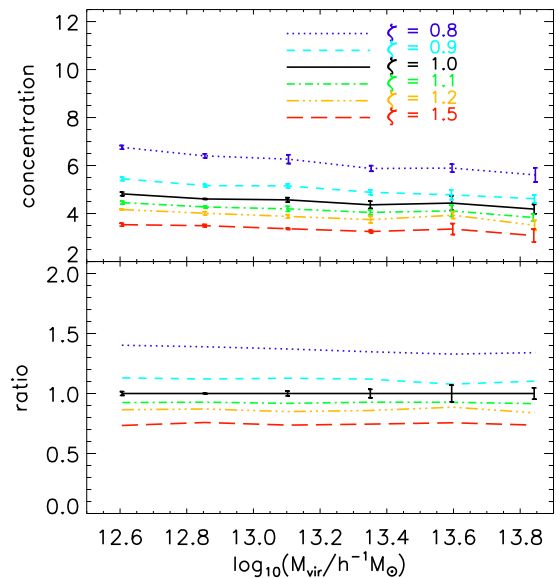}
\caption{Same  as Fig.~\ref{fig:concentration}  but for  the virial  halo mass
  $M_{\rm  vir}$ defined  by a  value  $\Delta_{\rm vir}$  from the  spherical
  collapse model.  }
\label{fig:concen_vir}
\end{figure}

In this paper we have compared  the halo mass function, concentration and bias
based on a series of modified gravity N-body simulations. The modified gravity
model is  characterized by  a single parameter  $\zeta$, which  determines the
enhancement of particle acceleration with  respect to GR.  All the simulations
for different $\zeta$ values are  started from the same initial condition. The
redshifts  for different  $\zeta$ of  the  comparison of  halo properties  are
selected  when  they have  the  same  linear  power spectrum  as  $\Lambda$CDM
simulations at $z_S=1.2$.

\begin{figure}
\includegraphics[width=0.45\textwidth]{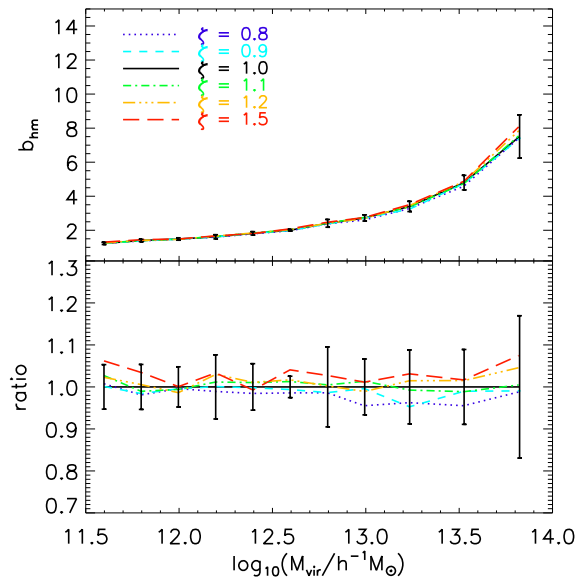}
\caption{Same  as Fig.~\ref{fig:bias}  but for  the virial  halo  mass $M_{\rm
    vir}$ defined  by a value  $\Delta_{\rm vir}$ from the  spherical collapse
  model.  }
\label{fig:bias_vir}
\end{figure}

Based on our various comparisons, we summarize our findings as follows.

\begin{itemize}

\item The measured halo mass functions can differ from that in GR by about
  $5-10\%$ for $10\%$ 
  deviation  from   GR ($|\zeta-1|=0.1$).  Although the difference must be
  symmetric with respect to $\zeta=1$ in the limit $\zeta\rightarrow 1$, it
  already shows significant  asymmetry for $|\zeta-1|\sim 10\%$.  For example,
  the difference between the halo mass  function of $\zeta=0.9$  and that of
  GR is about twice of the corresponding case of $\zeta=1.1$.  

\item For a  $10\%$ deviation from GR($|\zeta-1|=0.1$), the  deviation of halo
  concentration can differ by $10-20\%$, which shows that modified gravity
  can  significantly affect  the structure  formation on  small scale  and the
  internal properties of the dark matter halos.

\item  The halo bias is less sensitive to the nature of gravity. For all the
  $\zeta\ne1$ values used in our simulation ($0.8\leq \zeta\leq 1.5$),  the
  halo  bias  differ from that 
  of GR by  less  than $10\%$. 

\end{itemize}

We note that all  the above findings are obtained based on  the halos that are
defined to  be the  spherical over-density regions  within which  the average
over-density are  $\Delta_{200}=200$ times the mean  density $\rho_{\rm mean}$
of the universe.  This definition is widely used in literature  and is easy to
be implemented  both in simulations and  in observations. The  impact of using
different   $\Delta$  has   been  discussed   using  high   resolution  N-body
simulations\citep{Tinker2008}.

One might argue that the differences between MG models and the GR shown above
are mainly  due to the  definition of  the halos. To  test to what  extent halo
definition  impacts  our results, we  have also compared  the halo
mass function, concentration and bias using the virial halo mass $M_{\rm vir}$
(defined  by a  value $\Delta_{\rm  vir}$ \citep{Eke1996}  from  the spherical
model) instead  of $M_{200}$. 
According to the virial mass definition from GR, we
do   see  in   Fig.~\ref{fig:ms_vir}  that   for  a   $20\%$   deviation  from
($|\zeta-1|<0.2$), the  deviation of halo  mass function differs by  less than
$10\%$, which is smaller than that of the halo mass $M_{200}$. However, we can
still detect the observed differences of the halo mass functions for different
$\zeta$ values.

In Fig.~\ref{fig:concen_vir},  for $\zeta>1.0$, the difference  of $c_{\rm
  vir}(\zeta)$ with respect to that of GR is   roughly  the  same  as  the case of
$c_{200}(\zeta)$.  Since at redshift  $z_S  =1.2$, the universe is dominated by
dark matter and $R_{200}$ is approximately 
equal to $R_{\rm vir}$, which is $195.7$ for our adopted parameters. 
In  one word, by comparing
the      results      shown      in      Fig.~\ref{fig:concentration}      and
Fig.~\ref{fig:concen_vir},  we  can  see   that  there  exists  a  significant
difference  of the  halo  concentration for  different  $\zeta$ values,  which
illustrates that  deviations from general relativity can strongly  affect the
mass distribution in the nonlinear regime, although the large-scale structures
are remarkably similar (see Fig.~\ref{fig:slice}) \footnote{This behavior
  should  also hold for more realistic modified gravity models with screening
  mechanisms required to pass the solar system tests. In this case, the halo
  center may be well shielded. Nevertheless, the accretion history and the
  halo outskirts are impacted by modifications to GR.  }.

Fig.~\ref{fig:bias_vir} shows that the bias of halos defined with
virial mass still shows weak
dependence on gravity,  for all the $\zeta\ne1$ considered ($0.8\leq \zeta\leq
1.5$).  This confirmes previous finding that the halo bias may not be a sensitive probe of
gravity. This conclusion would be robust for halos of mass less than
$10^{13}\msun$.  Neverthless, our simulations do not have sufficiently large
box size to robustly 
measure the bias of halos with  mass larger  than $10^{13}\msun$ and
evaluate its dependence on gravity. 

In th above results $\Delta_{\rm vir}$ is evaluated assuming GR. It is known
that  $\Delta_{\rm vir}$ is  changed in  MG models \citep{Schmidt2009a, 
Schmidt2010}. To what extent it can account for the found difference in the mass
function \citep{Schmidt2009a, 
Schmidt2010} is an issue for further investigation.

Our  results  have  robustly demonstrated  that the  halo  mass
function  is  not completely determined by the shape and
amplitude of the linear power spectrum. As a consequence, it shows significant
dependence on the nature of gravity. We also show that the halo concentration
is also affected by gravity. These results can be used to understand the
impact of gravity on halo formation and distribution. They can also be used in
halo model to understand the nonlinear matter clustering \cite{Cui10}.

\acknowledgements

We would  like to thank Volker  Springel for the  N-GenIC, Postprocessing code
and detailed  help. We are grateful to the anonymous referee for useful 
comments that helped to improve the presentation of this paper. Youcai thanks  
Donghai Zhao for helpful  discussions. This work  is  supported  by  the  
national  science  foundation  of  China  (grant Nos.  10973027,   11025316, 
10973018, 10925314,  11233005,  11121062, 11203054), National  Basic Research 
Program of  China  (973 Program) under grant No.2009CB24901, the
CAS/SAFEA  International  Partnership  Program  for  Creative  Research  Teams
(KJCX2-YW-T23) and the Shanghai Committee of Science and Technology, China
(grant No. 12ZR1452800).

\end{document}